\documentclass[prd,aps,showpacs,amsmath,amssymb,twocolumn]{revtex4}
\usepackage{graphicx}
\usepackage{dcolumn}
\usepackage{bm}

\begin{document}

%

\let\a=\alpha      \let\b=\beta       \let\c=\chi        \let\d=\delta
\let\e=\varepsilon \let\f=\varphi     \let\g=\gamma      \let\h=\eta
\let\k=\kappa      \let\l=\lambda     \let\m=\mu
\let\o=\omega      \let\r=\varrho     \let\s=\sigma
\let\t=\tau        \let\th=\vartheta  \let\y=\upsilon    \let\x=\xi
\let\z=\zeta       \let\io=\iota      \let\vp=\varpi     \let\ro=\rho
\let\ph=\phi       \let\ep=\epsilon   \let\te=\theta
\let\n=\nu
\let\D=\Delta   \let\F=\Phi    \let\G=\Gamma  \let\L=\Lambda
\let\O=\Omega   \let\P=\Pi     \let\Ps=\Psi   \let\Si=\Sigma
\let\Th=\Theta  \let\X=\Xi     \let\Y=\Upsilon

%

%

\def\cA{{\cal A}}                \def\cB{{\cal B}}
\def\cC{{\cal C}}                \def\cD{{\cal D}}
\def\cE{{\cal E}}                \def\cF{{\cal F}}
\def\cG{{\cal G}}                \def\cH{{\cal H}}
\def\cI{{\cal I}}                \def\cJ{{\cal J}}
\def\cK{{\cal K}}                \def\cL{{\cal L}}
\def\cM{{\cal M}}                \def\cN{{\cal N}}
\def\cO{{\cal O}}                \def\cP{{\cal P}}
\def\cQ{{\cal Q}}                \def\cR{{\cal R}}
\def\cS{{\cal S}}                \def\cT{{\cal T}}
\def\cU{{\cal U}}                \def\cV{{\cal V}}
\def\cW{{\cal W}}                \def\cX{{\cal X}}
\def\cY{{\cal Y}}                \def\cZ{{\cal Z}}

\def\dbd{{$0\nu 2\beta\,$}}
\def\etabys{{$\frac{\eta}{s}$}}
%

\newcommand{\Ns}{N\hspace{-4.7mm}\not\hspace{2.7mm}}
\newcommand{\qs}{q\hspace{-3.7mm}\not\hspace{3.4mm}}
\newcommand{\ps}{p\hspace{-3.3mm}\not\hspace{1.2mm}}
\newcommand{\ks}{k\hspace{-3.3mm}\not\hspace{1.2mm}}
\newcommand{\des}{\partial\hspace{-4.mm}\not\hspace{2.5mm}}
\newcommand{\desco}{D\hspace{-4mm}\not\hspace{2mm}}



\title{\boldmath $\eta/s$ in a strongly coupled QFT}

\author{Namit Mahajan
}
\email{nmahajan@prl.res.in}
\affiliation{
 Theoretical Physics Division, Physical Research Laboratory, Navrangpura, Ahmedabad
380 009, India
}


\begin{abstract}
We consider $O(N)$ $g\varphi^4$ theory with the coupling $g$ being large, and calculate
shear viscosity to entropy density ratio ($\eta/s$). The final result for $\eta/s$ has a form
remarkably similar to that obtained from string theory calculations via the AdS/CFT conjecture.
The method adopted can be used to compute quantities of interest in other theories as well with some
modifications and reveals some very interesting features within the considered theory.
\end{abstract}

\pacs{
}
\maketitle


Study and characterization of various properties of fluids is perhaps one of the oldest subjects
in physics. One of the properties characterizing the goodness of a fluid is the shear viscosity, $\eta$.
In fact, often, inverse of shear viscosity, called fluidity, is used to quantify this aspect:
good fluids have small $\eta$. In a microscopic
theory, one would like to write the energy momentum tensor in terms of the field variables and then
compute various transport coefficients, like shear and bulk viscosities, conductivity etc using, say,
an effective kinetic theory description or employing Kubo formulae 
(see for example \cite{Hosoya:1983id}-\cite{Arnold:1998cy}). For weakly
coupled theories at high temperature, one has (throughout we'll be
working with natural units: $\hbar=c=k_B=1$):
\begin{equation}
 \eta = \kappa \frac{T^3}{\alpha^2 \ln(\alpha^{-1})}
\end{equation}
where $\kappa$ is a dimensionless parameter depending on the details of the theory and $\alpha$ is 
the appropriate coupling constant entering the amplitude: $\lambda$ in a scalar field theory,
$g^2/(4\pi)$ in a gauge theory. At such high temperatures,
the entropy density, $s$ varies as $T^3$. Therefore, both $\eta$ and $\eta/s$
are expected to be rather large. A fluid described by such a weakly coupled theory, therefore, can not
be expected to be a perfect fluid. The essential underlying physics is that the ratio, $\eta/s$,
is the ratio between the relaxation time $\tau$ and the typical time scale involved
in the quantum theory $t_q$. The latter is set by the inverse of temperature while
$\tau$ is governed by inverse of the scattering rate, which is proportional to square of the coupling.
Alternatively, $\eta/s$ is proportional to the ratio of
the mean free path of the particles to their average separation. A large mean free path, as
is there in a weakly coupled theory, implies that the momentum can be transported to
distances much larger than the average spacing between the particles.

In heavy ion collisions, extreme conditions are reached. Quarks and gluons are deconfined
and exist in a state referred to as quark gluon plasma (QGP). Under such extreme conditions, the
temperatures reached are still not very large that the QCD coupling is very small. Still, 
a moderately large $\eta/s \sim 1-2$ is obtained by interpolating the results at very high and low
temperatures \cite{Csernai:2006zz}.
Experiments however confirm that the QGP produced has a rather small $\eta/s$ 
$\sim {\mathcal{O}}(0.1-0.2)$, thereby
making it one of the most ideal or perfect fluids known \cite{Adler:2003kt}. For a comparison: $\eta/s\vert_{water} \sim 8$
and $\eta/s\vert_{He} \sim 1-2$. The experimental results thus pose a challenging problem. The produced QGP
is then to be described by a strongly coupled gauge theory rather than a weakly coupled one. In fact,
small value of $\eta/s$ has been the attributed reason behind the success of hydrodynamical description
of the heavy ion collision data assuming ideal fluid (or close to it) behaviour (see for example \cite{Kolb:2003dz}). 

One is now faced with a severe difficulty. Most of the known methods of computing quantities in a given
microscopic theory rely on perturbative expansion in terms of the coupling while
here one has a situation where no such method can be ever expected to give the desired results. The weak 
coupling result does not change much even after including the next order corrections or summing a set
of diagrams or adopting other methods (see \cite{Carrington:1999bw}). 
What is needed is a strong coupling calculation, for which no known methods seem to work.
In fact, lattice QCD calculations are also fraught with large uncertainties when attempting to
extract $\eta$ \cite{Aarts:2002cc}. However, in pure gluodynamics, remarkably accurate results have been obtained yielding
$\eta/s \sim {\mathcal{O}}(0.1)$ \cite{Meyer:2007ic}. Hopefully, in near future, similar high precision results will
become available for QCD.

While lattice QCD may provide accurate results in future, one may wonder if there is 
any way to even estimate quantities such as $\eta/s$ in strongly coupled theories. Such 
a machinery is provided by AdS/CFT correspondence \cite{Maldacena:1997re} (see also \cite{Gubser:1998bc}). 
In simple terms, it is a statement of
equivalence of four dimensional ${\mathcal{N}}=4$ Super Yang-Mills (SYM), a conformal field theory (CFT),
at strong coupling with a weakly coupled string theory defined on a ten dimensional space
AdS$_5\times$S$_5$. Operationally, on the gauge theory side, one has parameters:
gauge coupling, $g_{YM}$ and $N_c$, number of colours. In the limit $N_c\to \infty$ such that
with $g_{YM}$ small, 't Hooft coupling \cite{'tHooft:1973jz} $\lambda =g_{YM}^2 N_c$ is large, the string theory
effectively reduces to a supergravity (SUGRA) theory and the conjectured correspondence is then
about partition functions of the two theories being the same when evaluated
on the boundary of AdS$_5$. In most cases, S$_5$ does not play an important role and the
correspondence, also called gauge/gravity duality at times, can be expressed as:
\begin{eqnarray}
 Z_{string}[\Phi]_{\partial AdS} &\simeq& e^{-S_{SUGRA}} \nonumber \\
 &=& Z_{CFT}[\phi] = \langle exp(\int dx\, {\mathcal{O}}\phi)\rangle
\end{eqnarray}
where $\Phi$ denotes a generic field in AdS$_5$ space-time while $\phi$ is its value on
the four dimensional boundary denoted by $\partial AdS$. $\phi$ acts as a source for a gauge invariant
local operator, ${\mathcal{O}}$, on the boundary, and thus various correlation functions can be computed.
The main advantage and calculational power brought in by this correspondence is that there exists
a clear dictionary that maps gauge invariant operators on the CFT side on to the fields on the gravity
side. Since the theory of gravity in higher dimension is a weakly coupled one, at the leading order,
one has to solve for classical equation of motion at the boundary to obtain the correlation functions
in the strongly coupled gauge theory. To study gauge theories at finite temperature, one studies
black hole solutions in the gravity theory and the duality or the correspondence enables evaluating
the relevant correlation functions. The conjecture has passed a wide variety of tests when applied to 
${\mathcal{N}}=4$ SYM \cite{Muck:1998rr}. 
Relevant to the present context, employing the strategy suggested
by the correspondence, $\eta/s$ has been computed for ${\mathcal{N}}=4$ SYM for $N_c\to \infty$ \cite{Kovtun:2004de}
(see also \cite{Das:1996we}) and
the result turns out to be very simple and elegant, and is referred to as 
Kovtun-Son-Starinets (KSS) bound in literature:
\begin{equation}
 \frac{\eta}{s}\vert_{KSS} = \frac{1}{4\pi}
\end{equation}
The most striking feature of this result is that it is independent of the value of the coupling constant.
Sub-leading corrections have also been computed and they bring positive
contribution to the leading term. It has been then conjectured that this is a universal lower bound on 
$\eta/s$ and all known materials or systems obey this bound. Interestingly, there can be situations when this bound
can be violated \cite{Cherman:2007fj} but in any case, the consistent calculation yields a lower bound. 
Related to this, it has been asked if quantum field theory imposes a lower bound on $\eta/s$ \cite{Cohen:2007qr} . 
The answer is in negation and there are examples when this ratio can be tuned to zero or arbitrarily small 
values using external experimental parameters.

What is remarkable about the above result for $\eta/s$ is that the QGP produced in heavy ion collisions,
and also some cold atom systems respect the bound. This is quite a striking result since
at the basic level ${\mathcal{N}}=4$ SYM and say QCD are very different theories. For example,
QCD is not conformal while ${\mathcal{N}}=4$ SYM is conformal and unlike QCD does not confine. 
${\mathcal{N}}=4$ SYM being supersymmetric, with $N_c=3$, has far more degrees of freedom compared to QCD.
Also, the matter in both the theories has different representation: fundamental in QCD and adjoint
in ${\mathcal{N}}=4$ SYM. It is then rather surprising why QCD should have $\eta/s$, extracted 
by experiments, similar to ${\mathcal{N}}=4$ SYM. The reason lies in the fact that certain quantities
like $\eta/s$ do not depend on the number of degrees of freedom and also depend rather weakly on
other details of the theory. Motivated by the success of AdS/CFT correspondence in calculating
$\eta/s$, similar techniques have been applied to study various other systems and many properties
in situations when usual weak coupling perturbative methods fail. These include hadron dynamics and spectrum
in QCD, application to other non-perturbative quantities in heavy ion collisions, application
to cold atoms and condensed matter systems (see for example \cite{Peeters:2007ab}).

Although it has been a successful endevour employing the conjectured duality and
computing various quantities, in the end when various limits and approximations
are made to go towards the realistic situation, it is a priori not
clear if such a computation actually makes sense and how much can the results be trusted,
and what are the corrections when departing from the strict limits.
It would be nice to have an independent computation, of say $\eta/s$, in a strongly coupled theory
employing some field theoretic techniques. If such a computation yields results
which bear some resemblance to the AdS/CFT results, it may hint towards a deeper reason. On the 
other hand it is equally likely that the field theoretic results have no similarity with the
results obtained using the duality. In either case, there will be something very important to learn.
In what follows, we try to attempt such a computation within a scalar field theory at large
coupling without any recourse to AdS/CFT methods. We hasten to add that the final outcome
is a pleasant surprise and there is perhaps something important to be uncovered.

Let us first consider a scalar field theory with quartic coupling $g \varphi^4$ with the coupling
constant $g>0$ and very large. The (Euclidean) action for this theory, including the source term, is:
\begin{equation}
S[J] = \int d^4x \left[\underbrace{-\frac{1}{2}(\partial\varphi)^2 - \frac{1}{2}m^2\varphi^2}_{S_1} \underbrace{- g\varphi^4}_{S_2} - J\varphi\right]
\end{equation}
Before we delve into attempting a calculation at strong coupling, let us quickly recall the usual
perturbative approach when the coupling constant $g$ is a small parameter. In the standard approach,
the quadratic part of the action, the free field part, which we have deliberately denoted as $S_1$ 
(rather than the usual notation $S_0$) is solved and the effect of interaction, $S_2$ (usually
denoted as $S_{int}$) is introduced
by making a series expansion and evaluating the correlation functions or Green functions
to the order (expansion in $g$) desirable, and possible to compute. This whole approach works since
the basic integrals now become gaussian integrals and are exactly solved.
This procedure is efficiently encoded
in a set of Feynman rules which define the basic two point function, the propagator, and interaction vertices.
In case when $g$ is not a small parameter, such an approach fails to work. In what follows, we assume
$g\to \infty$. In such a situation, the usual perturbative methods are bound to be meaningless. But we may
try to invert the line of reasoning: let us attempt to first solve for the interaction part i.e. start with
$S_2$ and treat the quadratic
part, $S_1$, as a perturbation. We adopt a path integral approach wherein typical Green functions will be evaluated as
\begin{eqnarray}
G^{(n)}(x_1,....x_n) &=& \frac{\int{\mathcal{D}}\varphi\,
\varphi(x_1)....\varphi(x_n)e^{-S}}{\int{\mathcal{D}}\varphi\,e^{-S}} \\
&=& \frac{1}{{\mathcal{K}}}\,\prod_{i=1}^n\left(\frac{\delta}{\delta J(x_i)}\right) 
\int{\mathcal{D}}\varphi\,e^{-S[J]}\vert_{J=0} \nonumber
\end{eqnarray}
where  the normalization factor is given by
\begin{equation}
{\mathcal{K}} = \int{\mathcal{D}}\varphi\,e^{-S[J=0]} = 2\,\Gamma(5/4)\,\frac{1}{g^{1/4}}
\end{equation}
and $S[J] = \int d^4x [-(g\varphi^4 + J\varphi)]$. The numerator in Eq.(5) is easily evaluated and can be
expressed as a combination of Hypergeometric functions, which is then differentiated w.r.t. the source $J$
desired number of times and then set $J=0$ to get the relevant correlation functions. Since the aim is to
evaluate $\eta/s$, we follow the
kinetic theory approach which means that scattering cross-sections $2\to n$ are needed. The amplitude for $2\to 2$
scattering i.e. 4-point Green function works out to be
\begin{equation}
G^{(4)} = \frac{1}{4g} = {\mathcal{A}}
\end{equation}
while higher point Green functions are suppressed by higher powers of $g$. Therefore, for $g\to \infty$, it suffices to
consider only $2\to 2$ scattering. The above expression for the scattering amplitude implies that the scattering cross-section
vanishes in the infinitely strong coupling domain and therefore the shear viscosity will be very large in such a case. This
result is rather counter-intuitive as one would have expected a diametrically opposite behaviour compared to the weak
coupling case. To gain some understanding, consider taking the non-relativistic limit of the above theory. 
The quartic term, the so called 
two body term in the non-relativistic language, describes the scattering amplitude with a potential
$V(x,y) \sim g \delta^3(x-y)$. Such
a potential is singularly short ranged and repulsive. Therefore the two particles interacting
with each other via such a repulsive
$\delta$ function potential with infinite strength can not perceive each other. 
Such a behaviour is expected to hold in the relativistic case
as well, and therefore the above result. 

Next, let us consider $O(N)$ (with $N\to\infty$) symmetric version of the $\varphi^4$ interaction. The relevant terms
read $S_{2,N} = -\int d^4x\sum_{a=1}^{N}[g(\varphi^a\varphi^a)^2 + J^a\varphi^a]$. 
Once again we look at a typical $2\to 2$ scattering
amplitude as above. Employing the $N$-dimensional spherical symmetry,
the expressions in the path integral are written in 'polar coordinates'. 
The integrals now depend on the variable $N$ as well, and they again appear as
combinations of Hypergeometric functions. To proceed, we first expand these around $g=\infty$ and then make
another expansion around $N=\infty$. In the intermediate stages, we introduce UV cut-off $\Lambda$ and 
at various places also encounter combinations of the form $N+k$, where $k$ are integer values like $1,2,...10$ e.t.c. 
Since in the adopted approximation, $N\to\infty$, all these are simply replaced by $N$. After
long but straightforward algebra one obtains:
\begin{equation}
 {\mathcal{K}}_N = -\frac{\sqrt{\pi}\Lambda^N}{2N}
\end{equation}
and the numerator 
takes the form:
\begin{eqnarray}
\int{\mathcal{D}}\varphi\,
\varphi(x_1)....\varphi(x_4)e^{-S_N} &=& -\frac{\sqrt{\pi}\Lambda^N}{N} \Bigg[\frac{1}{4}\left(2-\sqrt{\frac{1}{g}}\right) 
 \nonumber \\
 &+& O\left(\frac{1}{N^2}\right)\Bigg]
\end{eqnarray}
Therefore, the scattering amplitude is thus given by
\begin{eqnarray}
 {\mathcal{A}}_N = \frac{1}{2} + O\left(\frac{1}{g^{1/2}}\right) + O\left(\frac{1}{N}\right) +
 O\left(\frac{1}{Ng^{1/2}}\right) + ...
\end{eqnarray}
This is the central result of the paper: the scattering amplitude for a strongly coupled large $N$ theory has been evaluated
at the leading order and turns out to be independent of the coupling constant. A heuristic way to
appreciate this result is that as seen above for a one component scalar field theory, the amplitude goes 
as inverse power of coupling. For $N$ components, the $N\to\infty$ limit compensates for this 
suppression, eventually yielding a non-zero answer.

As expected on physical grounds, the $\Lambda$ dependence exactly cancels without having
to discard any term by hand. We have explicitly checked that interchanging the order of $N$ expansion
and intermediate integrals also yields exactly the result. In performing the calculation
in this order, one encounters 'Hurwitz's LerchPhi' function. After taking the appropriate limits,
once again the $\Lambda$ dependence cancels and the end result is the same. This is reassuring as
a somewhat different method during the intermediate steps produces the same result.

We now turn to the evaluation of the shear viscosity. For this purpose, we adopt
the effective kinetic theory approach with a further simplification of assuming an
average relaxation time which is independent of energy. This last assumption is only
for simplifying the discussion and does not really affect the results apart from bringing
in some extra sub-leading terms. Following \cite{Jeon:1994if} and setting the mass
to zero (various integrals simplify in this limit and yield a simple exact form) we obtain
\begin{equation}
 \eta = \frac{\pi}{15} \frac{N\,T^3}{{\mathcal{A}}_N^2}
\end{equation}
while the entropy density is 
\begin{equation}
 s = \frac{2\pi^2}{45}N\,T^3
\end{equation}
We thereby have the shear viscosity to entropy density as
\begin{eqnarray}
 \frac{\eta}{s}\vert_N &=& \frac{1}{4\pi}\left(\frac{3}{2}\right) + O\left(\frac{1}{\sqrt{g}},\frac{1}{N}\right)
\end{eqnarray}
This is a rather surprising outcome. The result has the form exactly as that of the KSS result apart from
a minor ${\mathcal{O}}(1)$ multiplicative factor. It may be worthwhile to mention that the value
of this multiplicative factor can change a bit depending upon the approximations made at different steps
of the calculation. Apart from this minor ambiguity, the result is quite robust. What is really
surprising about the above result is that no where in the whole computation was any reference made
to the conjectured duality/correspondence but the final result bears a close
resemblance with the KSS result. 

The other approach towards evaluating transport coefficients, including $\eta$, is to employ Kubo's
relations which in this case imply looking at the two point function of the stress tensor. In the 
present context, one would only retain the $g\varphi^4$ term in the energy-momentum
tensor, $T_{\mu\nu}$. Each four point quartic vertex would contribute a factor of $1/g$ while
each of the two internal lines, which eventually will be cut, bring a factor of $N$ due to $N$ possible
ways to contract. The complete calculation is rather difficult at this stage as a very different form of
perturbation theory suited to this approach needs to be developed and then employed to compute a 
physical quantity \cite{namit}. What the above arguments indicate is that in the limit $g,N\to\infty$, the
dependence on $g$ would be compensated by dependence on $N$ and one hopes to get a result very similar to
obtained with the effective kinetic theory approach. However, we must confess that at this stage this 
is only an indication rather than a concrete statement.

Let us also say that a similar result is expected for a pure gauge theory at large coupling strength and 
large number of colours. In that case,
neglecting the cubic interaction vertex and retaining only the quartic gauge boson vertex,
and quantizing in a physical gauge, the structure is
quite similar to that of a multi-component scalar theory we have studied. It should be possible
to include the fermions as well though at this stage it is not clear if further approximations
will be needed to handle the fermion determinant. This is left for future study.

Having obtained $\eta/s$ in a strongly coupled $O(N)$ scalar theory employing a very non-standard 
path integral approach and the fact that the result looks like that obtained from AdS/CFT correspondence, one
wonders if this is a shear coincidence or there is something deeper about it. 
Also, it has been shown that gauge invariant operators
in the free field theory produce correlation functions that map on to AdS correlation functions \cite{Gopakumar:2003ns}.
Let us recall that
the path integral after only retaining the quartic term above has the form similar to matrix models,
and letting $g,N \to\,\infty$ could be like some double scaling limit 
(but not the usual double scaling limit that is commonly referred to) of such a matrix
model (see \cite{Bessis:1980ss}). It may be worthwhile to point out that the large $N$ limit considered here is
quite different compared to the usual one \cite{'tHooft:1973jz} where $g\to 0$ and $N\to\infty$ such that
$\lambda=g^2N$ is a large parameter. So what we have called a strongly coupled theory
is somewhat different in spirit compared to the usual one where it is $\lambda$ which is large. 
The calculation that has been performed is a genuine strong coupling calculation and the
similarity with the string theory calculations is rather tantalizing.
All these hints are somewhat speculative at this stage but the close similarity may be hinting at
something that is there to be uncovered. 

Until now we have completely neglected the quadratic part of the action and also set mass to zero
while evaluating the kinetic theory integrals. All this also implies that the theory in consideration
is scale invariant, and therefore the bulk viscosity, $\zeta$, vanishes. The ratio of the two viscosities, 
$\zeta/\eta$ acquires a different form in a strongly coupled theory compared to a weak coupling regime.
Systematically including the quadratic terms into the calculation would require developing a new set of Feynman rules
where the usual rules of the standard perturbation theory will not apply. This would however
determine various quantities including the ratio $\zeta/\eta$ in a self consistent manner. This is beyond
the scope of the present work and is left for a separate study.

In summary, we have attempted a calculation of computing
$\eta/s$ in a strongly coupled field theory without any reference to gauge/gravity duality. The
calculation is based on treating the qudratic part as a small perturbation (at the leading order
the quadratic part is simply ignored) and evaluating the exact path integral for the interaction part, which
is large since we assume the coupling $g\to\infty$. To the best of our knowledge, this is the first
calculation of this kind. For a single component $g\varphi^4$ with $g>0$, the scattering rate tends to zero,
implying that the shear viscosity in the infinitely large coupling regime will be very large like
the weakly coupled theory. This is rather counter-intuitive. When the same method is applied to
a multi-component theory with $N$ being simultaneously large,
the final outcome is a result remarkably similar to the AdS/CFT calculation of $\eta/s$. We thus speculate that
there is a lot more to learn about the strongly coupled theories and there are perhaps many more surprises
like the result above. We again remark that the ${\mathcal{O}}(1)$ difference between the result
above and AdS/CFT result could change depending upon the intermediate approximations and manipulations.
But the fact that the result, at the leading order, is independent of the coupling will hold.
It'll be interesting to extend this method to other theories
like Yukawa theory and to appropriate cases where the coupling, generically denoted as $g$, takes 
negative values. Another application could be in the context of self-interacting dark matter where
it could be strongly interacting but still yield a small cross-section. Such a method, perhaps, has
lot more potential and should be developed further in a systematic fashion.

\end{document}